\documentstyle[12pt]{article}
\def\zid{1\kern-0.36em\llap~1}

\newcommand{\beq}{\begin{equation}}
\newcommand{\ber}{\begin{eqnarray}}
\newcommand{\eeq}{\end{equation}}
\newcommand{\eer}{\end{eqnarray}}

\begin{document}

\begin{titlepage}
%\vbox {\vspace{0.1mm}} %Leaves space at top of 1st page.
\rightline{[SUNY BING 6/25/00] } \rightline{ hep-ph/0006342}
\vspace{2mm}
\begin{center}
{\bf TESTING FOR NEW COUPLINGS IN TOP QUARK
DECAY}\footnote{Contributed Paper for ICHEP2000; More detailed
paper is hep-ph/0007086.}\\ \vspace{2mm} Charles A.
Nelson\footnote{Electronic address: cnelson @ binghamton.edu  }
and L. J. Adler, Jr.\\ {\it Department of Physics, State
University of New York at Binghamton\\ Binghamton, N.Y.
13902-6016}\\[2mm]
\end{center}

%\vspace{2mm}

\begin{abstract}
Tests of the Lorentz structure of $t \rightarrow W^+ b $ decay
will be carried out at the Tevatron, and later at the LHC and at a
NLC.  To quantitatively assay future measurements of competing
observables, we consider the $g_{V-A}$ coupling values of the
helicity decay parameters versus \newline
          `` $(V-A)$ $+$ Single Additional Lorentz Structures ".
Three phase-type ambiguities exist, but measurement of the sign of
the large interference between the $W$ longitudinal/transverse
\newline amplitudes could exclude the two due dynamically to additional
$(S+P)$ and $(f_M + f_E)$ \newline couplings.  Sizable $
T$-violation signatures can occur for low-effective mass scales (
$ < 320 GeV $), but in most cases can be more simply excluded by $
10\% $ precision measurement of the \newline probabilities $
P(W_L)$ and $ P(b_L)$ . Signatures for the presence of $
T$-violation associated with the dynamical phase-type ambiguities,
$ CP$-violation signatures, and $ \Lambda_b $ polarimetry are also
discussed.

\end{abstract}

\end{titlepage}

\section{ Motivations, and Content Versus Ref.[3] }

In physics at the highest available energies, it is always
important to exploit simple reactions and decays so as to search
for new forces, for new dynamics, and for discrete symmetry
violations. Because the t-quark weakly decays before hadronization
effects are significant, and because of the large t-quark mass,
t-quark decay can be an extremely useful tool for such fundamental
searches. Initial tests of the Lorentz structure and of symmetry
properties of $t \rightarrow W^+ b $ decay will be carried out at
the Tevatron[1], but the more precise measurements will be
possible at the CERN LHC [2] and at a NLC [2].

It is important to be able to quantitatively assay future
measurements of competing observables consistent with the standard
model (SM) prediction of only a $g_{V-A}$ coupling and only its
associated discrete symmetry violations. For this purpose, without
consideration of possible explicit $T$ violation, in Ref.[3] plots
were given of the values of the helicity parameters in terms of a
``$(V-A)$ $+$ Additional Lorentz Structure" versus effective-mass
scales for new physics, $\Lambda_i$, associated with each
additional Lorentz structure.

In this {\bf contributed paper}, to assay future measurements of
helicity parameters in regard to $T$ violation, the effects of
possible explicit $T$ violation are briefly reported. A more
detailed paper on this latter subject will soon be available. In
effective field theory, $\Lambda_i$, is the scale [4] at which new
particle thresholds or new dynamics are expected to occur;
$\Lambda_i$ can also be interpreted as a measure of a top quark
compositeness/condensate scale.  In measurement of some of the
helicity parameters, the LHC should be sensitive to $ \sim 3 $ \%
and the Tevatron in a Run 3 to perhaps the $ \sim 10 $ \% level
(``ideal statistical error levels") [5].

\section{ Consequences of Single Additional Lorentz Structures in Absence of Explicit $T$ Violation }

In this section, we briefly review the work reported in Ref. [3].
This published paper contains a more detailed discussion, useful
simple formulas relating the `` $\alpha, \beta, \gamma$ " relative
phases of Fig.1a  and the helicity parameters of Fig.1b, and plots
of the values of the associated helicity parameters in the case of
single additional Lorentz structures.

The attached Figs. 1a, 1b provide a good orientation to this
topic: a complete measurement of on-shell properties of the $t
\rightarrow W^+ b $ decay mode will have been accomplished when
the 4 moduli are determined and any 3 of the relative phases of
the helicity
amplitudes $%
A(\lambda _{W^{+}},\lambda _b)$.  The helicity parameters appear
directly in various polarization and spin-correlation functions
such as those obtained in Ref.[5].

The top lines of the first two tables list the standard
model(SM)'s numerical values for the quantities shown in Figs. 1a,
1b.  In the SM, all the relative phases are either zero or $\pm
\pi$ so the primed helicity parameters are zero. In Table 1 in the
top line are the standard model expectations for the numerical
values of the helicity amplitudes $ A\left( \lambda_{W^{+} }
,\lambda_b \right) $ for $ t\rightarrow W^{+}b $ decay in $ g_L =
1 $ units. The input values are $m_t=175GeV, \; m_W = 80.35GeV, \;
m_b = 4.5GeV$. The $\lambda_b = 1/2$ b-quark helicity amplitudes
would vanish if $m_b$ were zero. For this reason, if one is guided
by the SM expectations, the most accessible quantities
experimentally should be the two moduli and the relative phase
shown on the right of Fig. 1a.  If the SM is correct, one expects
that the $A(0, -1/2)$ and $A(-1,-1/2)$ moduli and relative phase
$\beta_L$ will be the first quantities to be determined . The
$\lambda_b = 1/2$ moduli are factors of 30 and 100 smaller in the
SM. Interference measurements between the two columns are of order
${\cal O}(LR)$.  $L$ and $R$ denote the $b $ quark's helicity
$\lambda _b=\mp 1/2$.  Throughout this moduli-phase analysis of
top decays, intrinsic and relative signs of the helicity
amplitudes are specified in accordance with the standard
Jacob-Wick phase convention.

In Table 2 in the top line are the SM's numerical values of the
associated helicity parameters. Explicit formulas for the standard
model helicity amplitudes and for experimental distributions in
terms of these helicity parameters are given in Ref.[5].

The layout of the corners in  Fig. 1 has been chosen to reflect
the layout in the probability plots for $P(W_L)$ versus $P(b_L)$,
see Ref.[3] and Figs.5-6 below.  The quantities $$
\begin{array}{c}
P(W_L)= \mbox{Probability} \; W^{+} \; \mbox{ is longitudinally
polarized,} \; \lambda_{W^{+}}=0 \\ \hspace*{-26mm} P(b_L)= \;
\mbox{Probability} \; b \; \mbox{ is left-handed,} \;
\lambda_b=-1/2
\end{array}
$$ In terms of the first two helicity parameters of Table 2,
$P(W_L)= \frac{1+\sigma}{2} = 0.705(SM)$ and  $P(b_L)=
\frac{1+\xi}{2} = 1.00 (SM)$. So in the standard model, the
emitted $W$ boson should be $70 \%$ longitudinally polarized and
the emitted b-quark should be almost completely left-handed
polarized.

The ``arrows'' in the upper part of Fig. 1 define the measurable
$\alpha ,\beta ,\gamma $ relative phases between the four
amplitudes. For instance,
\begin{equation}
\alpha _0=\phi _0^R-\phi _0^L,\;\beta _L=\phi _{-1}^L- \phi
_0^L,\;\gamma _{+}=\phi _1^R-\phi _0^L\;
\end{equation}
where $A(\lambda _{W^{+}},\lambda _b)=|A|\exp (i\phi _{\lambda
_{W^{+}}}^{L,R})$. So for a pure $V-A$ coupling, the $\beta $'s
vanish and all the $\alpha $'s and $\gamma $'s equal $+\pi $ (or
$-\pi $) to give the intrinsic minus sign of the standard model's
$b_R$ amplitudes, see top row of Table 1.

The lower part of Fig. 1 displays the real part and imaginary part
(primed) helicity parameters corresponding to interference
measurements of the respective relative phases. For instance, c.f.
Appendix B of Ref.[3],
\begin{equation}
\begin{array}{c}
\eta _L\equiv \frac 1\Gamma |A(-1,-\frac 12)||A(0,-\frac 12)|\cos
\beta _L
\\
\eta _L^{\prime }\equiv \frac 1\Gamma |A(-1,-\frac 12)||A(0,-
\frac 12)|\sin \beta _L
\end{array}
\end{equation}
and
\begin{equation}
\eta _{L,R}=\frac 12(\eta \pm \omega )
\end{equation}

Because of the relative magnitudes of the moduli predicted by the
SM, in our consideration of information from b-quark polarization
measurements, we concentrate on the two b-quark interference
parameters $\kappa_0$ and $\epsilon_+$ and on their primed
analogues. If surprises are discovered in top quark decay, other
phases and/or helicity parameters might be more useful and
certainly would be useful as checks and/or constraints. By
$\Lambda _b$ polarimetry[5], or some other $b$-polarimetry
technique, it would be important to measure the $\alpha $ and
$\gamma $ relative phase. In the standard model, the two helicity
parameters between the amplitudes with the largest moduli are
\begin{equation}
\begin{array}{c}
\kappa _0\equiv \frac 1\Gamma |A(0,\frac 12)||A(0,-\frac 12)|\cos
\alpha _0
\\
\epsilon _{+}\equiv \frac 1\Gamma |A(1,\frac 12)||A(0,- \frac
12)|\cos \gamma _{+}
\end{array}
\end{equation}
We refer to $\kappa _0,\epsilon _{+}$ as the ``$b$-polarimetry
interference parameters''. For ${\kappa _0}^{'},{\epsilon
_{+}}^{'}$, the sine function replaces the cosine function in
Eqs.(4). Unfortunately from the perspective of a complete
measurement of the four helicity amplitudes, the tree-level values
of $\kappa _0,\epsilon _{+}$ in the SM are only about $1\%$. See
the top line in both parts of Table 2. Two dimensional plots of
the type $(\epsilon _{+},\eta _L)$ and $(\kappa _0,\eta _L)$, and
of their primed counterparts, have the useful property that the
unitarity limit is a circle of radius $\frac 12$ centered on the
origin[3].

In the plots in Ref.[3] and below, the values of the helicity
parameters are given in terms of a \newline ``$(V-A)$ + Single
Additional Lorentz Structure''. Generically, in the case of no
explicit $T$ violation, we denote these additional couplings by
\begin{equation}
g_{Total} \equiv g_L+g_X \\
\end{equation}
$$ X= \left\{ \begin{array}{ll} X_c =  \; \mbox{chiral} =
\{V+A,S\pm P,f_M\pm f_E\} \\ X_{nc} =  \; \mbox{non-chiral} =
\{V,A,S,P,f_M,f_E\}. \\
\end{array}
\right. $$ For \hskip1em  $t \rightarrow W^+ b$, the most general
Lorentz coupling is $ W_\mu ^{*} J_{\bar b t}^\mu = W_\mu ^{*}\bar
u_{b}\left( p\right) \Gamma ^\mu u_t \left( k\right) $ where $k_t
=q_W +p_b $, and
\begin{eqnarray}
\Gamma _V^\mu =g_V\gamma ^\mu + \frac{f_M}{2\Lambda }\iota \sigma
^{\mu \nu }(k-p)_\nu + \frac{g_{S^{-}}}{2\Lambda }(k-p)^\mu
\nonumber \\ +\frac{g_S}{2\Lambda }(k+p)^\mu
+%
\frac{g_{T^{+}}}{2\Lambda }\iota \sigma ^{\mu \nu }(k+p)_\nu
\end{eqnarray}
\begin{eqnarray}
\Gamma _A^\mu =g_A\gamma ^\mu \gamma _5+ \frac{f_E}{2\Lambda
}\iota \sigma ^{\mu \nu }(k-p)_\nu \gamma _5 +
\frac{g_{P^{-}}}{2\Lambda }(k-p)^\mu \gamma _5  \nonumber \\
+\frac{g_P}{2\Lambda }%
(k+p)^\mu \gamma _5  +\frac{g_{T_5^{+}}}{2\Lambda }\iota \sigma
^{\mu \nu }(k+p)_\nu \gamma _5
\end{eqnarray}
For $g_L = 1$ units with $g_i = 1$, the nominal size of
$\Lambda_i$ is $\frac{m_t}{2} = 88GeV$, see below.

Lorentz equivalence theorems for these couplings are treated in
Appendix A of Ref.[3]. Explicit expressions for the $A(\lambda
_{W^{+}},\lambda _b)$ in the case of these additional Lorentz
structures are given in Ref. [5]. Other recent general analyses of
effects in  $t\rightarrow W^{+}b$ decay associated with new
physics arising from large effective- mass
scales $%
\Lambda _i$ are in Refs. [6-12]. Some work on higher order QCD and
EW corrections has been done in [13].

The partial width $\Gamma $ for $t\rightarrow W^{+}b$ is the
remaining and  very important moduli parameter for testing for
additional Lorentz structures.  Since $\Gamma $ sets the overall
scale, it cannot be well measured by spin-correlation techniques,
which better measure the ratios of moduli and relative phases, so
we consider $\Gamma $ separately; see also [14,15].  From the
perspective of possible additional Lorentz structures, measurement
of the partial width $\Gamma $ is an important constraint. In
particular, this provides a strong constraint on possible $V+A$
couplings in contrast to measurement of $P(W_L)$ which does
not[3].  $\Gamma $ provides a useful constraint for the
possibility of additional $V$ and $A$ couplings which are
appealing from the perspective of additional gauge-theoretic
structures.

\section{Moduli Parameters and Phase-Type Ambiguities}

Versus predictions based on the SM, two dynamical phase-type
ambiguities were found by investigation of the effects of a single
additional ``chiral'' coupling $g_i$ on the three moduli
parameters $\sigma =P(W_L)- P(W_T),\;\xi =P(b_L)-P(b_R),\; $and
$\zeta =\frac 1\Gamma (\Gamma _L^{b_L-b_R}-\Gamma _T^{b_L-b_R})$.

For an additional $S+P$ coupling with $\Lambda _{S+P}\sim
-34.5GeV$ the values of $(\sigma ,\xi ,\zeta )$ and also of the
partial width $\Gamma $ are about the same as the SM prediction,
see Table 2. This is the first dynamical ambiguity. Table 1 shows
that this ambiguity will also occur if the sign of the
$A_X(0,-\frac 12)$ amplitude for $g_L+g_X$ is taken to be opposite
to that of the SM's amplitude. An additional $S\pm P$ only effects
the longitudinal $W^{\pm}$ amplitudes and not the transverse
$\lambda _W=\mp 1$ ones. By requiring that
\begin{equation}
\frac{A_X(0,-\frac 12)}{A_X(-1,-\frac 12)}=- \;\frac{A_L(0,-
\frac 12)}{%
A_L(-1,-\frac 12)}
\end{equation}
for $X=S+P$, we obtain a simple formula%
\begin{equation}
\Lambda _{S+P}=- (\frac{g_{S+P}}{g_L})\frac{m_t \;
q_W}{2(E_W+q_W)} \sim - (\frac{g_{S+P}}{g_L}) \frac{m_t}{4} ( 1-
(\frac{m_W}{m_t})^2 ).
\end{equation}
It is important to regard these ambiguities from (i) the signs in
their $b_L$ amplitudes versus those for the SM and from (ii) the
tensorial character and $\Lambda$ value of the associated Lorentz
structure.

For an additional $f_M+f_E$ coupling with $\Lambda _{f_M+f_E}\sim
53GeV$ the values of $(\sigma ,\xi ,\zeta )$  are also about the
same as the SM prediction, see Table 2. This is the second
dynamical ambiguity. In this case, the partial width $\Gamma $ is
about half that of the SM due to destructive interference.  Table
1 shows that this ambiguity will also occur if the sign of the
$A_X(-1,-\frac 12)$ amplitude for $g_L+g_X$ is taken to be
opposite to that of the SM's amplitude. Again, from (8) for
$X=f_M+f_E$,
we obtain%
\begin{equation}
\Lambda _{f_M+f_E}=
(\frac{g_{f_M+f_E}}{g_L})\frac{m_tE_W}{2(E_W+q_W)} \sim
(\frac{g_{f_M+f_E}}{g_L}) \frac{m_t}{4} ( 1+(\frac{m_W}{m_t})^2 )
\end{equation}
since $\frac{m_b}{m_t}\frac{\sqrt{E_b-
q_W}}{\sqrt{%
E_b+q_W}}\sim 10^{-3}$.

Besides the $f_M+f_E$ construction of this second phase- type
ambiguity, it should be kept in mind that some other mechanism
might produce the relative sign change shown in Table 1, but
without also changing the absolute value of the $b_L$ amplitudes.
In this case the measurement of the partial width  $\Gamma $ would
not resolve this phase ambiguity.

From consideration of Table 1, a third (phase) ambiguity can be
constructed by making an arbitrary sign-flip in the $b_L$
amplitudes, so $A_X(\lambda _{W,}\lambda _b=-\frac
12)=-A_{V-A}(\lambda _{W,}\lambda _b=- \frac 12)$, with no
corresponding sign changes in the $b_R$ amplitudes.

Resolution of this third ambiguity, as well as determination of
two remaining independent relative phases ( e.g. $\alpha_0$ and
$\gamma_+$ ) necessary for a complete amplitude measurement of $t
\rightarrow W^+ b $ decay, will require direct empirical
information about the $b_R$-amplitudes.  One way would be from a
$\Lambda_b$ polarimetry measurement [5] of the $b$-polarimetry
interference parameters $\epsilon_+$ and $\kappa_0$.  Even at an
NLC, such measurements will be difficult unless certain non-SM
couplings occur.    In particular, here additional $S+P$ and $f_M
+ f_E$ couplings have negligible effects, but non-chiral couplings
like $V$ or $A$, $f_M$ or $f_E$ (for $\epsilon_+$), $S$ or $P$
(for $\kappa_0$) can produce large effects[3].

Since the helicity parameters appear directly in the various
polarization and spin-correlation functions, it is clearly more
model independent to simply measure them rather than to set limits
on an `` ad hoc" set of additional coupling constants.  The large
$m_b$ effects displayed in some of the plots in Ref.[3] explicitly
demonstrate this point.  In many cases, finite $m_b$ effects in
both $b_L$ and $b_R$ amplitudes lead to sizable `` oval shapes" as
the effective mass scale $ \Lambda_i$ varies. There do not exist
``Lorentz equivalence theorems" with-respect-to both $m_b$
dependence and a minimal set of couplings when $m_b$ is allowed to
vary.

In summary, in the absence of explicit $T$ violation, three
phase-type ambiguities versus the SM prediction exist: two
dynamical ones with low effective mass scales, $g_{V-A} + g_{S+P}$
with $\Lambda_{S+P} \sim  -35 GeV$ and $g_{V-A} + g_{f_M + f_E }$
with $\Lambda_{ f_M + f_E } \sim  53 GeV$, and a third due to an
arbitrary sign-flip in the $b_L$-amplitudes $A_X (\lambda_b =-1/2)
= -A_{V-A} ( \lambda_b = -1/2)$.  The two dynamical ambiguities
can be resolved by measurement of the sign of the large
interference between the $W$ longitudinal/transverse amplitudes.
Measurement of the sign of the $\eta_L$ helicity parameter will
determine the sign of $cos \beta_L $ where $ \beta_L $ is the
relative phase of the two $b_L$-amplitudes ( $\eta_L = \pm 0.46 $
where the upper sign is for the SM ).  Both from the perspective
of carefully testing the SM and that of searching for new physics,
we believe that it is very important that experiments measure both
this $W$ longitudinal/transverse interference parameter and its
associated $T$ violation parameter ${\eta_L}^{'}$.  The latter
parameter is very important in the following analysis in this
paper.

\section{Consequences of Explicit $T$ Violation}

To assay future measurements of helicity parameters in regard to
$T$ violation, the next five sets of figures, Figs. 2-6, are for
the case of a single additional pure-imaginary coupling, $ i g_i /
2 \Lambda_i $ or $ i g_i $, associated with a specific additional
Lorentz structure, $ i = S, P, S + P , \ldots $.

In the $t$ rest frame, the matrix element for $t \rightarrow W^{+}
b$ is \beq \langle \theta _1^t ,\phi _1^t ,\lambda _{W^{+} }
,\lambda _b |\frac 12,\lambda _1\rangle =D_{\lambda _1,\mu
}^{(1/2)*}(\phi _1^t ,\theta _1^t ,0)A\left( \lambda _{W^{+} }
,\lambda _b \right) \eeq where $\mu =\lambda _{W^{+} } -\lambda _b
$ in terms of the $W^+$ and $b$ helicities. $\lambda_1$ gives the
$t$ quark's spin component quantized along a $z_1^t $ axis, see
Fig.1 in 2nd paper in Ref.[5].  So, upon a boost back to the $(t
\bar{t} )$ center-of-mass frame, or to the $\bar{t}$ rest frame,
$\lambda_1$ also specifies the helicity of the t quark.  By
rotational invariance there are only two amplitudes $A(0, -1/2),
A(-1, -1/2) $ for $\lambda_b = 1/2$, and two with $\lambda
_{W^{+}} = 0,1$ for $\lambda_b = -1/2$.
  For the $CP$-conjugate
process, $\bar t \rightarrow W^{-} \bar b$, in the $\bar t$ rest
frame \beq \langle \theta _2^t ,\phi _2^t ,\lambda _{W^{-}
},\lambda _{\bar b}|\frac 12,\lambda _2\rangle =D_{\lambda _2,\bar
\mu }^{(1/2)*}(\phi _2^t ,\theta _2^t ,0)B\left( \lambda _{W^{-}
},\lambda _{\bar b}\right) \eeq with $\bar \mu =\lambda
_{W^{-}}-\lambda _{\bar b}$.

As shown in Table 3 a specific discrete symmetry implies a
specific relation among the associated helicity amplitudes.  In
the case of $T$ invariance, the helicity amplitudes must be purely
real. The $T$ invariance of Table 3 will be violated if either (i)
there is a fundamental violation of canonical $T$ invariance, or
(ii) there are absorptive final-state interactions. In the SM,
there are no such final-state interactions at the level of
sensitivities considered in the present analysis.  In our earlier
papers[5], we have kept this assumption of ``the absence of
final-state interactions" manifest by referring to the $T$
invariance of Table 3 as ``$ \tilde{T}_{FS} $ violation".

Barred parameters $ \bar{\xi}, \bar{\zeta}, \ldots $ have the
analogous definitions for the $CP$ conjugate process, $\bar t
\rightarrow W^- \bar b $. Therefore, any $ \bar{\xi} \neq \xi,
\bar{\zeta} \neq \zeta, \ldots $ $ \Longrightarrow $ CP is
violated.  That is, ``slashed parameters" $\not\xi \equiv \xi
-
\bar{\xi}$, \ldots, could be introduced to characterize and
quantify the degree of CP violation.  This should be regarded as a
test for the presence of a non-CKM-type CP violation because,
normally, a  CKM-phase will contribute equally at tree level to
both the $t \rightarrow W^+ b_L$ decay amplitudes and so a
CKM-phase will cancel out in the ratio of their moduli and in
their relative phase.  There are four tests for non-CKM-type CP
violation[5].

A recent review of $CP$-violation in t-quark physics is in [16].

\subsection{Additional $S \pm P, , f_M \pm f_E , S, P, f_M,$ or $f_E$ couplings}

The two plots displayed in Fig.2 are for dimensional couplings
with chiral $ S \pm P, f_M \pm f_E $ and non-chiral $ S, P, f_M,
f_E $ Lorentz structures. The upper plot displays the $
{\eta_L}^{'} $ helicity parameter versus the effective-mass scale
$ \Lambda_i $ with $g_i = 1 $ in $g_L = 1 $ units. The lower plot
displays the induced effect of the additional coupling on the
partial width for $ t\rightarrow W^{+}b $.  The standard model
limit is at the ``wings" where $ \vert  \Lambda_i  \vert
\rightarrow \infty $ for each additional dimensional coupling.

Fig.3 displays plots of the b-polarimetry interference parameters
${\epsilon_+}^{'}$ and $ {\kappa_0}^{'}$ versus $ \Lambda_i $ for
the case of a single additional  $ S, P, f_M, f_E $ and $ S \pm P,
f_M - f_E $ coupling:  Curves are omitted in the plots in this
paper when the couplings produce approximately zero deviations in
the helicity parameter of interest, e.g. this occurs for $ f_M +
f_E $ in both the ${\epsilon_+}^{'}$ and $ {\kappa_0}^{'}$
helicity parameters. The unitarity limit for ${\epsilon_+}^{'}$
and $ {\kappa_0}^{'}$ is also $ 0.5 $ .

\subsection{Additional $V+A, V,$ or $A$ couplings}

An additional $V-A$ type coupling with a complex phase versus the
SM's $g_L$ is equivalent to an additional overall complex factor
in the SM's helicity amplitudes. This will effect the overall
partial width $\Gamma$, but it can't otherwise be observed by
spin-correlation measurements.

For a single additional gauge-type coupling $V, A, $ or $V+A$, in
Fig.4 are plots of the b-polarimetry interference parameters
${\epsilon_+}^{'}$ and $ {\kappa_0}^{'}$, and of the partial width
for $ t\rightarrow W^{+}b $ versus pure-imaginary coupling
constant $ i g_i $.  The $g_i $ value is in $g_L = 1 $ units.  In
the cases of the additional dimensionless, gauge-type couplings,
the standard model limit is at the origin, $g_i \rightarrow 0 $.

\subsection{Indirect effects of $T$ violation on other helicity parameters}

The plots in Fig.5 show the indirect effects of a single
additional pure-imaginary chiral coupling, $ i g_i / 2 \Lambda_i $
or $ i g_i $, on other helicity parameters.  For the coupling
strength ranges listed in the ``middle table", the upper plot
shows the effects on the probability, $P(W_L)$, that the emitted $
W^{+} $ is ``Longitudinally" polarized and the effects on the
probability, $P(b_L)$, that the emitted b-quark has ``Left-handed"
helicity. Each curve is parametrized by the magnitude of the
associated $ g_i$ or $ \Lambda_i$.  On each curve, the central
open circle corresponds to the region with a maximum direct $T$
violation signature, e.g. for $ f_M + f_E $ from Fig. 2 this is at
$ \vert \Lambda_{f_M + f_E} \vert \sim 50 GeV $.  The large/small
solid circles correspond respectively to the ends of the ranges
listed in the middle table where the direct signatures fall to
about $ 50\%$ of their maximum values. Similarly the lower plot is
for the W-polarimetry interference parameters $ \eta, \omega$.
Curves are omitted for the remaining moduli parameter $ \zeta$
since a single additional pure-imaginary coupling in these ranges
produces approximately zero deviations from the pure $V-A$ value
of $\zeta = 0.41 $.

The plots in Fig.6 show the indirect effects of a single
additional pure-imaginary non-chiral coupling on other helicity
parameters. Versus the middle table given here, the curves are
labeled as in Fig. 5. The upper plot is for the two probabilities
$P(W_L)$ and $P(b_L)$. The lower plot is for the W-polarimetry
interference parameters $ \eta, \omega$.

In summary, sizable $ T$-violation signatures can occur for
low-effective mass scales ( $ < 320 GeV $) as a consequence of
pure-imaginary couplings associated with a specific additional
Lorentz structure.  However, in most cases, such additional
couplings can be more simply excluded by $ 10\% $ precision
measurement of the probabilities $ P(W_L)$ and $ P(b_L)$.  The
W-polarimetry interference parameters $\eta$ and $\omega$ can also
be used as indirect tests, or to exclude such additional
couplings.

\section{Tests for $T$ Violation Associated with the Dynamical Phase-Type Ambiguities}

In Fig.7 are plots of the signatures for a partially-hidden $T$
violation associated with a $S+P$ phase-type ambiguity: We require
Eq.(8) to hold when the additional $S+P$ coupling, $g_{S+P} / 2
\Lambda_{S+P} $ has a complex effective mass scale parameter $
\Lambda_{S+P} = \vert \Lambda_{S+P} \vert \exp{ - i \theta} $
where $\theta$ varies with the mass scale $\vert \Lambda_{S+P}
\vert$. For $m_b = 0$, the resulting function $\theta ( \vert
\Lambda_{S+P} \vert ) $ is very simple. This construction
maintains the standard model values in the massless b-quark limit
for the four moduli parameters, $ P(W_L), P(b_L) , \zeta,$ and
$\Gamma$. The function $\theta ( \vert \Lambda_{S+P} \vert ) $ is
then used for the $S+P$ coupling when $ m_b = 4.5 GeV $. The SM
values for the moduli parameters are essentially unchanged. There
are two cases, $ \sin{ \theta} \geq 0 $ and $ \sin{\theta } \leq 0
$. The phase choice of ${\phi^R}_1 = \pm \pi$, cf. top line in
Table 1, has no consequence since it is a $ 2 \pi$ phase
difference.

For $ \sin{ \theta} \geq 0 $ in Fig.7 is the solid curve for the $
{\eta_L}^{'} $, the $T$ violation W-polarimetry interference
parameter, plotted versus $ 1 / \vert \Lambda_{S+P} \vert$.  The
dashed curve is for the W-polarimetry interference parameters
$\eta_L, \eta, \omega$ which are degenerate. The dark rectangles
show the standard model values at the $\vert \Lambda_{S+P}
\vert\rightarrow \infty $ endpoint where $\theta = \pi / 2$. At
the other endpoint $\vert \Lambda_{S+P} \vert \sim 34.5 GeV $, or
$ 1 / \vert \Lambda_{S+P} \vert = 0.029 GeV^{-1}$, the coupling is
purely real with $\theta = \pi$. The unitarity limit for each of
these helicity parameters is $0.5$.

From the perspectives of (i) measuring the $W$ interference
parameters and of (ii) excluding this type of $T$ violation, it is
noteworthy that where $ {\eta_L}^{'} $ has the maximum deviation,
there is a zero in $\eta_L, \eta, \omega$.  So if the latter
parameters were found to be smaller than expected or with the
opposite sign than expected, this would be consistent with this
type of $T$ violation.

At the maximum of $ {\eta_L}^{'} $, $ \vert \Lambda_{S+P} \vert
\sim 49 GeV$ and the other $T$ violation parameters are also
maximum. The curves for these parameters have the same over all
shape as  $ {\eta_L}^{'} $ but their maxima are small,
${\epsilon_+}^{'} \sim 0.015$ and ${\kappa_0}^{'} \sim 0.028$.

For the other case where $ \sin{\theta } \leq 0 $, all these $T$
violation primed parameters have the opposite overall sign.  The
signs of other helicity parameters are not changed.

In Fig.8 are plots of the signatures for a partially-hidden $T$
violation associated with a $f_M+f_E$ phase-type ambiguity: As
above for the analogous $S+P$ construction, the additional
$f_M+f_E$ coupling $g_{f_M+f_E} / 2 \Lambda_{f_M+f_E} $ now has an
effective mass scale parameter $ \Lambda_{f_M+f_E} = \vert
\Lambda_{f_M+f_E} \vert \exp{ - i \theta} $ in which $\theta$
varies with the mass scale $\vert \Lambda_{f_M+f_E} \vert$ to
maintain standard model values in the massless b-quark limit for
the moduli parameters $ P(W_L), P(b_L) ,$ and $ \zeta$. For the
case $ \sin{ \theta} \geq 0 $,  in Fig.8 the upper plot shows by
the solid curve the $T$ violation W-polarimetry interference
parameter $ {\eta_L}^{'} $ versus $ 1 / \vert \Lambda_{f_M+f_E}
\vert$.  By the dashed curve, it shows the W-polarimetry
interference parameters $\eta_L, \eta, \omega$ which are
degenerate.  At the endpoint $\vert \Lambda_{f_M+f_E} \vert \sim
52.9 GeV $, or $ 1 / \vert \Lambda_{f_M+f_E} \vert = 0.0189
GeV^{-1}$, the coupling is purely real with $\theta = 0$.

Here, as in Fig. 7, where $ {\eta_L}^{'} $ has the maximum
deviation, there is a zero in $\eta_L, \eta, \omega$. The lower
plot shows the indirect effect of such a coupling on the partial
width $\Gamma$ for  $ t\rightarrow W^{+}b $.

At the maximum of $ {\eta_L}^{'} $, $ \vert \Lambda_{f_M+f_E}
\vert \sim 63 GeV$.  The curve for the $T$ violation parameter
${\kappa_0}^{'}$ has the same shape and is also maxmimum at the
same position with a value ${\kappa_0}^{'} \sim 0.005$.
${\epsilon_+}^{'} $ remains very small.  For the other case where
$ \sin{\theta } \leq 0 $, each of these $T$ violation primed
parameters has the opposite overall sign.

In summary, sufficiently precise measurement of the W-interference
parameter $\eta_L$ and of the $ {\eta_L}^{'} $ parameter can
exclude partially-hidden $T$ violation associated with either of
the two dynamical phase-type ambiguities. However, if $\eta_L = (
\eta + \omega ) / 2 $ were found to be smaller than expected or
with a negative sign, this would be consistent with this type of
$T$ violation.

\begin{center}
{\bf Acknowledgments}
\end{center}

For computer services, one of us (CAN) thanks John Hagan and Ted
Brewster. This work was partially supported by U.S. Dept. of
Energy Contract No. DE-FG 02-86ER40291.

\newpage

\begin{center}
{\bf Table Captions}
\end{center}

Table 1: For the
ambiguous moduli
points, numerical values of the associated helicity
amplitudes $ A\left( \lambda_{W^{+} } ,\lambda_b \right)
$.
The values for the amplitudes are listed first in $ g_L =
1 $
units, and second as $ A_{new} = A_{g_L = 1} / \surd
\Gamma $
which removes the effect of the differing partial width,
$
\Gamma $ for $ t\rightarrow W^{+}b $. [$m_t=175GeV, \;
m_W =
80.35GeV, \; m_b = 4.5GeV$ ].

Table 2:  For the ambiguous moduli points, numerical values of the
associated helicity parameters.  Listed first are the four moduli
parameters. Listed second are the values of the $W$-polarimetry
interference parameters which could be used to resolve these
dynamical ambiguities.

Table 3:  The helicity formalism is based on the assumption of
Lorentz invariance but not on any specific discrete symmetry
property of the fundamental amplitudes, or couplings.  For
instance, for $ t\rightarrow W^{+}b $ and $ \bar{t}\rightarrow
W^{-}\bar{b} $ a specific discrete symmetry implies a definite
symmetry relation among the associated helicity amplitudes.

\begin{center}
{\bf Figure Captions}
\end{center}

FIG. 1: For $ t\rightarrow W^{+}b $ decay, display of the four
helicity amplitudes $ A\left( \lambda _{W^{+} } ,\lambda _b
\right) $ relative to the $W^{+}$ boson and  b-quark helicities.
The {\bf upper sketch} defines the measurable `` $ \alpha, \beta,
\gamma $ " relative phases, c.f. Eqs(1). The {\bf lower sketch}
defines the real part and imaginary part (primed) helicity
parameters corresponding to these relative phases.  Measurement of
a non-zero primed helicity parameter would be a direct signature
for $T$ violation.

FIG. 2: To assay future measurements of helicity parameters in
regard to $T$ violation, the next {\bf five sets of figures} are
for the case of a single additional pure-imaginary coupling, $ i
g_i / 2 \Lambda_i $ or $ i g_i $, associated with a specific
additional Lorentz structure, $ i = S, P, S + P , \ldots $ .  The
two plots displayed here are for dimensional couplings with chiral
$ S \pm P, f_M \pm f_E $ and non-chiral $ S, P, f_M, f_E $ Lorentz
structures. The {\bf upper plot} displays the $ {\eta_L}^{'} $
helicity parameter versus the effective-mass scale $ \Lambda_i $
with $g_i = 1 $ in $g_L = 1 $ units. The {\bf lower plot} displays
the induced effect of the additional coupling on the partial width
for $ t\rightarrow W^{+}b $.  The standard model limit is at the
``wings" where $ \vert  \Lambda_i  \vert \rightarrow \infty $ for
each additional dimensional coupling. The unitary limit for $
{\eta_L}^{'} $ is $ 0.5 $.

FIG. 3: Plots of the b-polarimetry interference parameters
${\epsilon_+}^{'}$ and $ {\kappa_0}^{'}$ versus $  \Lambda_i $ for
the case of a single additional  $ S, P, f_M, f_E $ and $ S \pm P,
f_M - f_E $ coupling:  Curves are omitted in the plots in this
paper when the couplings produce approximately zero deviations in
the helicity parameter of interest, e.g. this occurs for $ f_M +
f_E $ in both the ${\epsilon_+}^{'}$ and $ {\kappa_0}^{'}$
helicity parameters. The unitarity limit for ${\epsilon_+}^{'}$
and $ {\kappa_0}^{'}$ is also $ 0.5 $ .

FIG. 4: For a single additional gauge-type coupling $V, A, $ or
$V+A$, plots of the b-polarimetry interference parameters
${\epsilon_+}^{'}$ and $ {\kappa_0}^{'}$, and of the partial width
for $ t\rightarrow W^{+}b $ versus pure-imaginary coupling
constant $ i g_i $.  The  $g_i $ value is in $g_L = 1 $ units.  In
the cases of the additional dimensionless, gauge-type couplings,
the standard model limit is at the origin, $g_i \rightarrow 0 $.

FIG. 5:  These plots show the indirect effects of a single
additional pure-imaginary chiral coupling, $ i g_i / 2 \Lambda_i $
or $ i g_i $, on other helicity parameters.  For the coupling
strength ranges listed in the ``middle table", the {\bf upper
plot} shows the effects on the probability, $P(W_L)$, that the
emitted $ W^{+} $ is ``Longitudinally" polarized and the effects
on the probability, $P(b_L)$, that the emitted b-quark has
``Left-handed" helicity.  Each curve is parametrized by the
magnitude of the associated $ g_i$ or $ \Lambda_i$.  On each
curve, the central open circle corresponds to the region with a
maximum direct $T$ violation signature, e.g. for $ f_M + f_E $
from Fig. 2 this is at $ \vert \Lambda_{f_M + f_E} \vert \sim 50
GeV $.  The large/small solid circles correspond respectively to
the ends of the ranges listed in the middle table where the direct
signatures fall to about $ 50\%$ of their maximum values.
Similarly the {\bf lower plot} is for the W-polarimetry
interference parameters $ \eta, \omega$. Curves are omitted for
the remaining moduli parameter $ \zeta$ since a single additional
pure-imaginary coupling in these ranges produces approximately
zero deviations from the pure $V-A$ value of $\zeta = 0.41 $.  A
dark rectangle denotes the value for the pure $V-A$ coupling of
the standard model.

FIG. 6:  These plots show the indirect effects of a single
additional pure-imaginary non-chiral coupling on other helicity
parameters.  Versus the middle table given here, the curves are
labeled as in Fig. 5. The {\bf upper plot} is for the two
probabilities $P(W_L)$ and $P(b_L)$. The {\bf lower plot} is for
the W-polarimetry interference parameters $ \eta, \omega$.

FIG. 7:  Plots of the signatures for a partially-hidden $T$
violation (see text) associated with a $S+P$ phase-type ambiguity:
In this case, the additional $S+P$ coupling, $g_{S+P} / 2
\Lambda_{S+P} $, has an effective mass scale parameter $
\Lambda_{S+P} = \vert \Lambda_{S+P} \vert  \exp{ - i \theta} $
where $\theta$ varies with the mass scale $\vert \Lambda_{S+P}
\vert$ to maintain standard model values in the massless b-quark
limit for the four moduli parameters, $ P(W_L), P(b_L) , \zeta,$
and $\Gamma$.  Plotted versus $ 1 / \vert \Lambda_{S+P} \vert$ for
the case $\sin {\theta} \geq 0$  is the solid curve for the $
{\eta_L}^{'} $, the $T$ violation W-polarimetry interference
parameter and the dashed curve for the W-polarimetry interference
parameters $\eta_L, \eta, \omega$ which are degenerate.  For $\sin
{\theta} \leq 0$, the $ {\eta_L}^{'} $ sign is opposite.  The dark
rectangles show the standard model values at the $\vert
\Lambda_{S+P} \vert \rightarrow \infty $ endpoint where $\theta =
\pi / 2$. At the other endpoint $\vert \Lambda_{S+P} \vert \sim
34.5 GeV $, or $ 1 / \vert \Lambda_{S+P} \vert = 0.029 GeV^{-1}$,
the coupling is purely real with $\theta = \pi$.  Where $
{\eta_L}^{'} $ has the maximum deviation, there is a zero in
$\eta_L, \eta, \omega$.

FIG. 8:  Plots of the signatures for a partially-hidden $T$
violation (see text) associated with a $f_M+f_E$ phase-type
ambiguity: The additional $f_M+f_E$ coupling, $g_{f_M+f_E} / 2
\Lambda_{f_M+f_E} $, has an effective mass scale parameter $
\Lambda_{f_M+f_E} = \vert \Lambda_{f_M+f_E} \vert  \exp{ - i
\theta} $ where $\theta$ varies with the mass scale $\vert
\Lambda_{f_M+f_E} \vert$ to maintain standard model values in the
massless b-quark limit for the moduli parameters $ P(W_L), P(b_L)
,$ and $ \zeta$. Versus $ 1 / \vert \Lambda_{f_M+f_E} \vert$ for
$\sin {\theta} \geq 0$, the {\bf upper plot} shows by the solid
curve $ {\eta_L}^{'} $.  By the dashed curve, it shows $\eta_L,
\eta, \omega$ which are degenerate.  At the endpoint $\vert
\Lambda_{f_M+f_E} \vert \sim 52.9 GeV $, or $ 1 / \vert
\Lambda_{f_M+f_E} \vert = 0.0189 GeV^{-1}$, the coupling is purely
real with $\theta = 0$. For $\sin {\theta} \leq 0$, the $
{\eta_L}^{'} $ sign is opposite.
 The {\bf
lower plot} shows the indirect effect of such a coupling on the
partial width $\Gamma$ for  $ t\rightarrow W^{+}b $.

\end{document}